\documentclass[showpacs,prb,aps,superscriptaddress,twocolumn,floatfix]{revtex4}
\usepackage{graphicx,float,amsmath,amssymb,color}
\newfont{\ensmathquatorze}{msbm10 scaled 1400}
\newfont{\ensmathonze}{msbm10 scaled 1100}
\newfont{\ensmathdix}{msbm10}
\newfont{\ensmathneuf}{msbm10 scaled 833}
\newfont{\ensmathhuit}{msbm10 scaled 694}
\newfam\ensmathfam
\textfont\ensmathfam=\ensmathonze
\scriptfont\ensmathfam=\ensmathdix
\scriptscriptfont\ensmathfam=\ensmathhuit

\newcommand{\ket}[1]{|\kern.3ex#1\kern.3ex\rangle}
\newcommand{\bra}[1]{\langle\kern.3ex #1 \kern.3ex|}

\newcommand{\EXP}[1]{{\mbox{\large e}}^{#1}}         

\newcommand{\re}{\mathop{\mathrm{Re}}\nolimits}      
\newcommand{\im}{\mathop{\mathrm{Im}}\nolimits}      

\def\I{{\rm i}}                  
\def\D{{\rm d}}                  

\newcommand{\drond}[2]{\frac{\partial #1}{\partial #2}} 

\newcommand\ab{{\alpha\beta}}

\begin{document}

\title{Al'tshuler-Aronov correction to the  conductivity of \\a large metallic square network}

\author{Christophe Texier}
\affiliation{Laboratoire de Physique Th\'eorique et Mod\`eles Statistiques, UMR 8626 du CNRS, Universit\'e Paris-Sud, F-91405 Orsay Cedex, France.}
\affiliation{Laboratoire de Physique des Solides, UMR 8502 du CNRS, Universit\'e Paris-Sud, F-91405 Orsay Cedex, France.}

\author{Gilles Montambaux}
\affiliation{Laboratoire de Physique des Solides, UMR 8502 du CNRS, Universit\'e Paris-Sud, F-91405 Orsay Cedex, France.}

\date{April 5, 2007}




\begin{abstract}
  We consider the correction $\Delta\sigma_\mathrm{ee}$ due to
  electron-electron interaction to the conductivity of a weakly disordered
  metal (Al'tshuler-Aronov correction). The correction is related to the
  spectral determinant of the Laplace operator. The case of a large square
  metallic network is considered. The variation of
  $\Delta\sigma_\mathrm{ee}(L_T)$ as a function of the thermal length $L_T$ is
  found very similar to the variation of the weak localization
  $\Delta\sigma_\mathrm{WL}(L_\varphi)$ as a function of the phase coherence
  length. Our result for $\Delta\sigma_\mathrm{ee}$ interpolates between the
  known 1d and 2d results, but the interaction parameter entering the
  expression of $\Delta\sigma_\mathrm{ee}$ keeps a 1d behaviour. Quite
  surprisingly, the result is very close to the 2d logarithmic behaviour
  already for $L_T\sim{a}/2$, where $a$ is the lattice parameter.
\end{abstract}

\maketitle

\section{Introduction}

At low temperature, the classical (Drude) conductivity of a weakly disordered
metal is affected by two kinds of quantum corrections~: the first one is the
{\it weak localization} (WL) correction, a phase coherent contribution that
originates from quantum interferences between reversed electronic
trajectories. This contribution to the averaged conductivity depends on the
phase coherence length $L_\varphi$ and the magnetic field~:
$\Delta\sigma_\mathrm{WL}(\mathcal{B},L_\varphi)$. The temperature manifests
itself through $L_\varphi$, since phase breaking may depend on temperature,
e.g. if it originates from electron-electron\cite{AltAroKhm82} or
electron-phonon\cite{ChaSch86} interaction.

In a metal, an electron is not only elastically scattered on the disordered
potential, but, due to the electron-electron interaction, is also scattered by
the electrostatic potential created by the other electrons. At low
temperatures, when the elastic scattering rate ($1/\tau_e$) dominates
the electron-electron scattering rate ($1/\tau_\mathrm{ee}(T)$), the motion of
the electron is diffusive between scattering events with other electrons.
In this regime,  electron-electron interaction is responsible for
a small depletion of the density of states at Fermi energy (called
the DoS anomaly or the Coulomb dip) and a correction to the
averaged conductivity as well, the so-called {\it
Al'tshuler-Aronov} (AA)
correction~\cite{AltAro79,AltKhmLarLee80,AltAro83,Fin83,CasCasLeeMa84,AltAro85,LeeRam85}
(see Refs.~\cite{AleAltGer99,AkkMon04,footnote1} for a recent discussion). 
AA and WL corrections are of the same order (but this latter vanishes in a
magnetic field). However, contrary to the WL, the AA correction is not
sensitive to phase coherence and involves another important length scale~: the
thermal length $L_T=\sqrt{D/T}$ ($\hbar=k_B=1$). The AA correction, denoted
below $\Delta\sigma_\mathrm{ee}(L_T)$, has been measured in metallic wires in
several
experiments~\cite{WhiTinSkoFla82,EchGerBozBogNil94,PieGouAntPotEstBir03,BauMalSchMaiEskSam05}.
From the experimental point of view, AA correction allows to study interaction
effects in weakly disordered metals, but also furnishes a local probe of
temperature in order to control Joule heating
effects~\cite{EchGerBozBogNil94,BauMalSchMaiEskSam05}, which is crucial in a
phase coherent experiment.

All the works aforementioned refer to the quasi-one-dimensional (wire) or
two-dimensional (plane) situations. Quantum transport has also been
studied in more complex geometries like networks of quasi-1d wires. For
example several studies of WL have been provided on large regular networks in
honeycomb and square metallic networks~\cite{PanChaRamGan85,DolLicBis86},
in square networks etched in a 2DEG~\cite{FerAngRowGueBouTexMonMai04}, 
and in square and dice silver
networks~\cite{SchMalMaiTexMonSamBau07}.  Theoretical studies of WL on
networks have been initiated by the works of Dou\c{c}ot \&
Rammal~\cite{DouRam85,DouRam86} and improved by Pascaud \&
Montambaux~\cite{PasMon99} who introduced a powerful tool\cite{footnote2}~:
the spectral determinant of the Laplace operator, that will be used in the
following (see also Ref.~\cite{AkkComDesMonTex00}).

The aim of this paper is to study how the AA correction can be
computed in networks. In a first part we briefly recall how the
spectral determinant can be used to compute the WL. Then in a
second part we will consider the AA correction.


\section{Spectral determinant and weak localization\label{sec:wl}}

\begin{figure}[!ht]
\begin{center}
\includegraphics[scale=0.4]{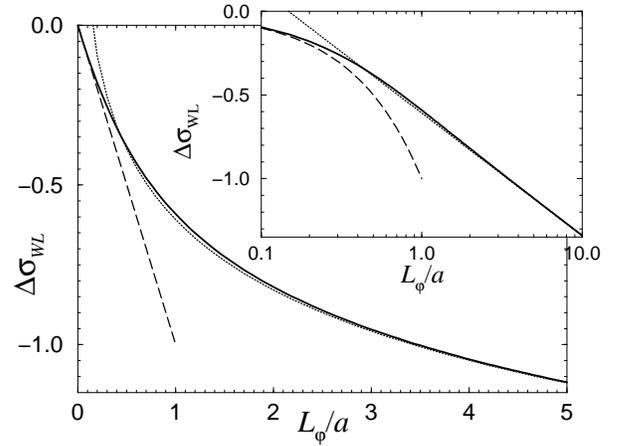}
\end{center}
\caption{\label{WLmesh}\it  
    $\Delta\sigma_\mathrm{WL}$ in unit of $2e^2/h$ as a function of $L_\varphi/a$
    (at zero magnetic field). 
    The dashed line is the 1d result. The dotted line is the 2d limit 
    eq.~(\ref{WLlimit2d}).}
\end{figure}

Interferences of reversed electronic trajectories are encoded in the Cooperon,
solution of a diffusion-like equation
$(\partial_t-D[\nabla-2\I{}eA(x)]^2)\mathcal{P}_c(x,x';t)=\delta(x-x')\delta(t)$,
where $A(x)$ is the vector potential. On large regular networks, when it is
justified to integrate uniformly the Cooperon over the network (see
Ref.~\cite{TexMon04} for a discussion of this point) it is meaningful to
introduce the space-averaged Cooperon 
$\mathcal{P}_c(t)=\int\frac{\D{x}}{\mathrm{Vol}}\mathcal{P}_c(x,x;t)$ then
\begin{eqnarray}
  \label{WL}
  \Delta\sigma_\mathrm{WL} &=&
  -\frac{2e^2  D}{\pi}
  \int_0^\infty\D t\,\EXP{-t/ \tau_\varphi}\,\mathcal{P}_c(t)
 \\
  \label{WL2}
  &=&
  -\frac{2e^2}{\pi}\frac1{\mathrm{Vol}}
  \frac{\partial}{\partial\gamma} \ln S(\gamma)
\end{eqnarray}
where  $\tau_\varphi=L_\varphi^2/D$ is the phase coherence time.
The factor $2$ stands for spin degeneracy.  
We have omitted in (\ref{WL},\ref{WL2}) a factor $1/s$ where $s$ is the
cross-section of the wires.
The parameter $\gamma$ is related to the phase coherence length
$\gamma=1/L_\varphi^2$ (note that description of the decoherence due to
electron-electron interaction in networks requires a more refined
discussion~\cite{LudMir04,TexMon05b}). The spectral determinant of
the Laplace operator is formally defined as
$S(\gamma)=\det(\gamma-\Delta)=\prod_n(\gamma+E_n)$ where $\{E_n\}$ is the
spectrum of $-\Delta$ [in the presence of a magnetic field,
$\Delta\to(\nabla-2\I eA)^2$]. The interest in introducing $S(\gamma)$ is that
it can be related to the determinant of a $V\times{V}$-matrix, where $V$ is
the number of vertices, that encodes all information on the network (topology,
length of the wires, magnetic field, connection to reservoirs). We label
vertices by greek letters. $l_\ab$ designates the length of the wire $(\ab)$
and $\theta_\ab$ the circulation of the vector potential along the wire. The
topology is encoded in the adjacency matrix~: $a_\ab=1$ if $\alpha$ and
$\beta$ are linked by a wire, $a_\ab=0$ otherwise. $\lambda_\alpha=\infty$ if
$\alpha$ is connected to a reservoir and $\lambda_\alpha=0$ if not. We
introduce the matrix
\begin{eqnarray}
  \mathcal{M}_\ab = \delta_\ab
  \left(
   \lambda_\alpha
   + \sqrt\gamma\sum_\mu a_{\alpha\mu}\coth\sqrt\gamma l_{\alpha\mu}
  \right)
  \nonumber\\
  -a_\ab \,\sqrt\gamma \frac{\EXP{-\I\theta_\ab}}{\sinh\sqrt\gamma l_\ab}
\end{eqnarray}
where the $a_{\alpha\mu}$ constraints the sum to run over neighbouring
vertices. Then~\cite{PasMon99}
\begin{eqnarray}
  S(\gamma) = \prod_{(\ab)}\frac{\sinh\sqrt\gamma l_\ab}{\sqrt\gamma}\:
  \det\mathcal{M}
\end{eqnarray}
where the product runs over all wires. We now consider a large square network
of size $N_x\times{N_y}$ made of wires of length $l_\ab=a$ $\forall(\ab)$. For
simplicity we impose periodic boundary conditions (topology of a torus), which
is inessential as soon as the total size of the network remains small compared
to $L_\varphi$.  At zero magnetic field the spectrum of the adjacency matrix
is $\epsilon_{n,m}=2\cos(2n\pi/N_x)+2\cos(2m\pi/N_y)$, with $n=1,\cdots,N_x$
and $m=1,\cdots,N_y$. Therefore
\begin{eqnarray}
&&
  S(\gamma)=\left(2\frac{\sinh\sqrt\gamma a}{\sqrt\gamma}\right)^{N_xN_y}
  \\\nonumber
&&
  \times
  \prod_{n=1}^{N_x}\prod_{m=1}^{N_y}
  \left(2\cosh\sqrt\gamma a - \cos\frac{2\pi n}{N_x}- \cos\frac{2\pi m}{N_y}
  \right)
  \:.
\end{eqnarray}
The calculation of $\ln{S(\gamma)}$ involves a sum that can be replaced by 
an integral when $N_x,\,N_y\gg{L_\varphi/a}$. Using 
\begin{equation}
  \label{RelUtile}
  \int_{0}^{2\pi}\frac{\D x\D y}{(2\pi)^2}\:
  \frac1{2A+\cos x+\cos y} = \frac1{\pi A}\,\mathrm{K}(1/A)
  \:,
\end{equation}
where $\mathrm{K}(x)$ is the complete elliptic integral of first
kind~\cite{gragra}, yields~\cite{FerAngRowGueBouTexMonMai04}
\begin{eqnarray}
\label{Ssqnet}
  \frac{1}{{\rm Vol}}\drond{}{\gamma}\ln S(\gamma)
  &=&\frac1{4\sqrt{\gamma}}
  \left[
    \coth\sqrt{\gamma}a - \frac{1}{\sqrt{\gamma}a}
  \right.
   \nonumber \\
&&\hspace{-1cm}
  \left.
    + \frac{2}{\pi}\tanh\sqrt{\gamma}a\:
    \mathrm{K}\left(\frac{1}{\cosh\sqrt{\gamma}a}\right)
  \right]
  \:,
\end{eqnarray}
where the volume of the network is ${\rm Vol}=2N_xN_ya$.  We recover the
expression of the WL first derived by Dou{\c c}ot \& Rammal \cite{DouRam86}.
Figure~\ref{WLmesh} displays the dependence of the WL correction as a function
of the phase coherence length $L_\varphi$. We now discuss two limiting cases.

\vspace{0.15cm}

\noindent{\it 1d limit.--}
In the limit $L_\varphi\ll{a}$ ({\it i.e.} $\sqrt{\gamma}a\gg1$)~:
\begin{equation}
  \label{WLlimit1d}
  \Delta\sigma_\mathrm{WL} = -\frac{2e^2}{h}\,
  \left( L_\varphi-\frac{L_\varphi^2}{2a}+O\left(\EXP{-2a/L_\varphi}\right) \right)
\end{equation}
We compare with the result for a wire of length $a$ connected at its extremities~:
$
\Delta\sigma_\mathrm{WL}^\mathrm{wire}
\simeq-\frac{2e^2}{h}\,(L_\varphi-\frac{L_\varphi^2}{a})
$.
As we can see the dominant terms coincide. Deviations appear when $L_\varphi/a$ 
increases since trajectories begin to feel the topology of the network.
This is already visible by comparing the second terms of the expansions.

\vspace{0.15cm}

\noindent{\it 2d limit.--}
In the limit $L_\varphi\gg{a}$ ({\it i.e.} $\sqrt{\gamma}a\ll1$),
we obtain
\begin{eqnarray}
  \frac{1}{{\rm Vol}}\drond{}{\gamma}\ln S(\gamma)=
  \frac{a}{2\pi}
  \bigg[
    \ln(4L_\varphi/a)+\frac{\pi}{6}
\nonumber\\
    +O\left(\frac{a^2}{L_\varphi^2}\ln\frac{L_\varphi}{a}\right)
  \bigg]
\end{eqnarray}
The conductivity reads
\begin{equation}
  \label{WLlimit2d}
  \Delta\sigma_\mathrm{WL}
  \simeq-\frac{2e^2}{h}\,a
  \left[\frac1{\pi}\ln(L_\varphi/a)+C_\mathrm{WL}\right]
\end{equation}
with $C_\mathrm{WL}=\frac{2\ln2}{\pi}+\frac16\simeq0.608$. As noticed in the
beginning of the section, eqs.~(\ref{WLlimit1d},\ref{WLlimit2d}) should be
divided by the cross-section $s$ of the wires. In the 2d limit, diffusive
trajectories expand over distances larger than $a$ and feel the two
dimensional nature of the system, being the reason why (\ref{WLlimit2d}) is
reminiscent of the 2d result. It is interesting to point that the network
provides a natural cutoff (the length of the wires, $a$) while the computation
of the WL for a plane in the diffusion approximation requires to introduce a
cutoff by hand for lower times in eq.~(\ref{WL}), which is the elastic
scattering time $\tau_e$. In this latter case the constant added to the
logarithmic behaviour is not well controlled since it depends on the cutoff
procedure (the computation of the constant for a plane requires to go beyond
the diffusion approximation and leads to~\cite{CasSha94}
$
\Delta\sigma^\mathrm{plane}_\mathrm{WL}
=-\frac{e^2}{\pi h}\ln(2L_\varphi^2/\ell_e^2+1)
\simeq-\frac{2e^2}{h}[\frac1{\pi}\ln(L_\varphi/\ell_e)+\frac1{2\pi}\ln2]
$ since $\ell_e\ll{L}_\varphi$).


\section{Al'tshuler-Aronov correction}

At first order in the electron-electron interaction, the exchange term is the
dominant contribution to the correction to the
conductivity~\cite{AltAro85,footnote3,AleAltGer99,AkkMon04}
\begin{eqnarray}
   \label{formula51}
   \Delta\sigma_\mathrm{ee} &=&
   -\frac{2\sigma_0}{d\pi\mathrm{Vol}}
   \int_{-\infty}^{+\infty}\D\omega\,
   \drond{}{\omega}\left(\omega\coth\frac\omega{2T}\right)
   \nonumber\\
   && \times
   \im\sum_{\vec q} D\vec q\,^2\,
   \frac{U(\vec q,\omega)}{(-\I\omega+D\vec q\,^2)^3}
\end{eqnarray}
where $U(\vec q,\omega)$ is the dynamically screened interaction. Within the
RPA approximation and in the small $\vec q$ and $\omega$ limit, the
interaction takes the form\cite{footnote4}
$U(\vec{q},\omega)\simeq\frac{1}{2\rho_0}\frac{-\I\omega+D\vec{q}\,^2}{D\vec{q}\,^2}$
where $\rho_0$ is the density of states per spin channel.  
Replacing the Drude conductivity by its expression $\sigma_0=2e^2\rho_0D$ and
performing an integration by parts, we get
\begin{eqnarray}
    \Delta\sigma_\mathrm{ee} &=& -\frac{2e^2D}{\pi d\mathrm{Vol}}
    \int\D\omega\,
     \drond{^2}{\omega^2}\left(\omega\coth\frac\omega{2T}\right)
    \nonumber\\
   && \times
\re\sum_{\vec q} \frac1{-\I\omega+D\vec q\,^2}
\end{eqnarray}
After Fourier transform, the result can be cast in the form~\cite{AkkMon04}~:
\begin{equation}
\label{sigmaee}
  \Delta\sigma_\mathrm{ee}=-\lambda_\sigma\frac{e^2  D}{\pi}
  \int_0^\infty\D t\,\left(\frac{\pi {T} t}{\sinh\pi {T} t}\right)^2
  \mathcal{P}_d(t) \ ,
\end{equation}
For the exchange term considered here, one finds $\lambda_\sigma= 4/d$.
Further calculation yields~\cite{AltAro85}
$\lambda_\sigma\simeq\frac4d-\frac32F$, where $F$ is the average of the
interaction on the Fermi surface (see definition in
Refs.~\cite{AltAro85,LeeRam85}). This expression of $\lambda_\sigma$ is valid
in the perturbative regime, $F\ll1$~; nonperturbative expression is given in
Refs.~\cite{Fin83,CasCasLeeMa84,AltAro85,LeeRam85}. $\mathcal{P}_d(t)$ is the
space integrated return probability
$\mathcal{P}_d(t)=\int\frac{\D{x}}{\mathrm{Vol}}\mathcal{P}_d(x,x;t)$, where
$\mathcal{P}_d(x,x';t)$ is solution of a classical diffusion equation similar
to the equation for $\mathcal{P}_c(x,x';t)$, apart that it does not feel the
magnetic field~:
$[\partial_t-D\Delta]\mathcal{P}_d(x,x';t)=\delta(x-x')\delta(t)$. Therefore
the Laplace transform of $\mathcal{P}_d(t)$ is given by
$\partial_\gamma\ln{S}(\gamma)$ with $\theta_\ab=0$. It is interesting to
point out that (\ref{sigmaee}) has a similar structure to (\ref{WL}) with a
different cutoff procedure for large time. It also involves a different
scale~: the temperature dependence of $\Delta\sigma_\mathrm{ee}$ is driven by
the length scale $L_T$ instead of $L_\varphi$ for the weak-localization
correction $\Delta\sigma_\mathrm{WL}$.

Up to eq.~(\ref{sigmaee}) the discussion is rather general and nothing has
been specified on the system. We have seen in section~\ref{sec:wl} that the WL
for the square network presents a dimensional crossover from 1d to 2d by
tuning $L_\varphi/a$. A similar dimensional crossover occurs for the AA
correction by tuning $L_T/a$ as we will see. This remark raises the question
of the dimension $d$ in eq.~(\ref{formula51}). To answer this question we
should return to the origin of the factor $1/d$~: the current lines in the 
conductivity $\sigma_{ij}$ produce a factor $q_iq_j$ replaced by
$\delta_{ij}\frac1d\vec{q}\,^2$ after angular integration. Since in a network
the diffusion in the wires has a 1d structure (provided that
$W\ll{L_T}\sim\sqrt{{D}/{\omega}}$, where $W$ is the width of the wires), the
dimension in $\lambda_\sigma$ is~$d=1$. Therefore we have for the
network~$\lambda^\mathrm{network}_\sigma\simeq4-\frac32F$.

If one now expands the thermal function in (\ref{sigmaee}) as~:
\begin{equation}
  \left(\frac{y}{\sinh y}\right)^2=4y^2\sum_{n=1}^\infty n\,\EXP{-2ny}
  \:,
\end{equation}
 we can also relate $\Delta \sigma_\mathrm{ee}$  to the spectral determinant. We obtain~:
\begin{equation}
  \label{mainresult}
  \Delta\sigma_\mathrm{ee}=-\lambda_\sigma\frac{e^2}{\pi{\rm Vol}}
  \sum_{n=1}^\infty \frac1n
  \left[
      \gamma^2\drond{^3}{\gamma^3}\ln S(\gamma)
  \right]_{\gamma=\frac{2n\pi}{L_T^2}}
\end{equation}
which is the central result of this paper. It is the starting point of the
discussion below.

\vspace{0.15cm}

\begin{figure}[!ht]
\begin{center}
\includegraphics[scale=0.35]{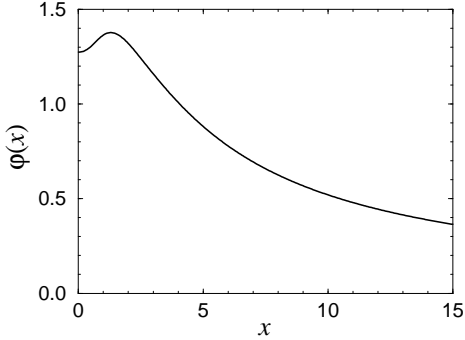}
\end{center}
\caption{\label{seem_f}\it The function $\varphi(x)$ of eq.~(\ref{defphi}).}
\end{figure}

\begin{figure}[!ht]
\begin{center}
\includegraphics[scale=0.4]{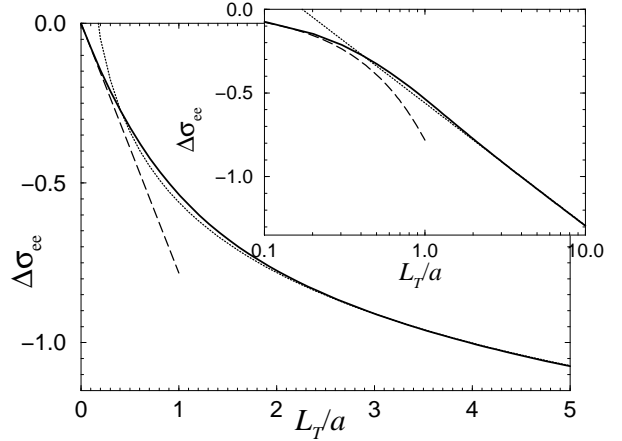}
\end{center}
\caption{\label{sigeemesh}\it
         The continuous line is
         $\Delta\sigma_\mathrm{ee}$ in unit of $\lambda_\sigma\frac{e^2}{h}$ 
         as a function of $L_T/a$ (the series (\ref{seeseries}) is computed 
         numerically). 
         The dashed line
         is the 1d limit, eq.~(\ref{see1dlimit}), and the
         dotted curve is the 2d limit, eq.~(\ref{see2dlimit}).}
\end{figure}

\noindent{\it Application to the case of the square network.--}
We have to compute $\gamma^2\drond{^3}{\gamma^3}\ln S(\gamma)$.
We start from (\ref{Ssqnet})  and compute its second derivative.
We obtain after some algebra~:
\begin{equation}
  \label{seeseries}
  \Delta\sigma_\mathrm{ee}=-\lambda_\sigma\frac{e^2}{h}\, \frac{a}{8}\:
  \sum_{n=1}^\infty \frac1n\,\varphi\left(\sqrt{2n\pi}\frac{a}{L_T}\right)
\end{equation}
where  the function $\varphi(x)$ is given by~:
\begin{widetext}
\begin{eqnarray}
\label{defphi}
&&\varphi(x)=
  -\frac{8}{x^2} + \frac{2x\cosh x}{\sinh^3x} + \frac{3}{\sinh^2x}
  +\frac{3\coth x}{x}
\nonumber\\
&&\hspace{0.75cm}
  +\frac2\pi
  \left\{
    \left[\frac{3\tanh x}{x}-3\right]\mathrm{K}\left(\frac{1}{\cosh x}\right)
    +\left[3-\frac{2x}{\sinh2x}\right]\mathrm{E}\left(\frac{1}{\cosh x}\right)
  \right\}
  \:,
\end{eqnarray}
\end{widetext}
$\mathrm{E}(x)$ being the complete elliptic integral of second
kind~\cite{gragra}.  The function $\varphi(x)$ is plotted in
figure~\ref{seem_f} and its limiting behaviours are easily
obtained~\cite{gragra}~:
\begin{eqnarray}
  \varphi(x)&=&\frac4\pi+O(x^2) \hspace{1.45cm}     \mbox{ for } x\to0 \\
  \label{phiexp2}
            &=&\frac{6}{x}-\frac8{x^2}+O(x\EXP{-2x})\mbox{ for } x\to\infty
  \hspace{0.5cm}
\end{eqnarray}
The $L_T$ dependence of AA correction on a square network is displayed on
figure~\ref{sigeemesh}, where we have plotted $\Delta\sigma_\mathrm{ee}(L_T)$
given by eq.~(\ref{seeseries}).  The dimensional crossover now occurs by
tuning the ratio~$L_T/a$. We consider the two limits.

\vspace{0.15cm}

\vspace{0.15cm}

\noindent{\it 1d limit.--}
For $L_T\ll{a}$ we can replace the expansion (\ref{phiexp2}) in the series
(\ref{seeseries}). Therefore
\begin{equation}
  \label{see1dlimit}
  \Delta\sigma_\mathrm{ee}\simeq-\lambda_\sigma\frac{e^2}{h}\,
  \left( 
    \frac{3\zeta(3/2)}{4\sqrt{2\pi}}\,L_T
    -\frac\pi{12} \frac{L_T^2}{a}
  \right)
\end{equation}
with $\frac{3\zeta(3/2)}{4\sqrt{2\pi}}\simeq0.782$.
The dominant term again coincides with the one for a connected
wire~\cite{AltAro85,AleAltGer99,AkkMon04} while the second differs by a factor $2$, as 
for the~WL [see discussion after eq.~(\ref{WLlimit1d})].

\vspace{0.15cm}

\noindent{\it 2d limit.--}
In the limit $L_T\gg{a}$ we introduce $\mathcal{N}=(L_T/a)^2$ and cut the sum
(\ref{seeseries}) in two pieces~:
$\sum_1^\infty=\sum_1^\mathcal{N}+\sum_\mathcal{N}^\infty$.  It is clear from
the limit behaviours of $\varphi(x)$ that the first sum diverges
logarithmically with $\mathcal{N}$ while the second brings a negligible
contribution of order $\mathcal{N}^0$. Therefore~:
\begin{equation}
  \label{see2dlimit}
  \Delta\sigma_\mathrm{ee} \simeq -\lambda_\sigma \frac{e^2}{h}\,
  a\left[\frac{1}{\pi}\ln(L_T/a) +C_\mathrm{ee}\right]
\end{equation}
The constant is estimated numerically. We find $C_\mathrm{ee}\simeq0.56$.

The two eqs.~(\ref{see1dlimit},\ref{see2dlimit}) should be divided by the
cross-section $s$ of the wires.

The two functions $\Delta\sigma_\mathrm{WL}(\mathcal{B}=0,L_\varphi)$
(figure~\ref{WLmesh}) and $\Delta\sigma_\mathrm{ee}(L_T)$
(figure~\ref{sigeemesh}) are very similar. Apart from the prefactors $2e^2/h$
and $\lambda_\sigma{e}^2/h$ which account respectively for the spin degeneracy
and the interaction strength, the linear behaviours at the origin have a
different slope ($1$ and $0.782$) and the logarithmic behaviours are slightly
shifted~: $C_\mathrm{WL}\simeq0.61$ and~$C_\mathrm{ee}\simeq0.56$.


%


\section{Comparison with experiments\label{sec:experiment}}

The AA correction has been recently measured by Mallet {\it et
  al}~\cite{MalSamBau07} in networks of silver wires with $3\:10^4$ and $10^5$
cells, lattice spacing $a=0.64\:\mu$m and diffusion constant
$D\simeq100\:$cm$^2/$s. The diffusion constant $D$ has been measured
separately (through measurement of the Drude conductance), therefore we can
compare our result (\ref{seeseries}) with experiment using one fitting
parameter only~: the interaction parameter~$\lambda_\sigma$. The 2d
logarithmic behaviour (\ref{see2dlimit}) has been observed in the range
$100\:$mK$<T<1\:$K from which the value
$\lambda^\mathrm{exp}_\sigma\simeq3.1$ was extracted, in agreement with
similar measurements performed on a long silver wire for
which~\cite{SamMohWebDegBau07,MalSamBau07}
$\lambda^\mathrm{exp,\:wire}_\sigma\simeq3.2$. We now compare with the
theoretical value~: for silver Fermi wavelength is $k_F^{-1}=0.083\:$nm and
Thomas-Fermi screening length $\kappa^{-1}=1/\sqrt{8\pi\rho_0e^2}=0.055\:$nm.
In the Thomas-Fermi approximation, the parameter $F$ is given
by\cite{AkkMon04} $F=(\frac{\kappa}{2k_F})^2\ln[1+(\frac{2k_F}{\kappa})^2]$,
therefore $F\simeq0.58$. Using the 1d nonperturbative
expression~\cite{AltAro85}
$\lambda_\sigma=4+\frac{48}{F}(\sqrt{1+F/2}-1-F/4)$, we get
$\lambda^\mathrm{th}_\sigma\simeq3.24$, close to the experimental value.


\section{Conclusion}

Equations (\ref{mainresult},\ref{seeseries}) are our main results. 
The first one shows that AA and WL can be formally related~:
\begin{eqnarray}
  \label{relationGilles}
  \Delta\sigma_\mathrm{ee}(L_T)
  &=&\frac{\lambda_\sigma}{2}\,
  \\ \nonumber
  && \hspace{-1cm}\times 
  \sum_{n=1}^\infty\frac1n
  \left[
   \gamma^2\drond{^2}{\gamma^2}\Delta\sigma_\mathrm{WL}(L_\varphi)
  \right]_{\gamma\equiv\frac1{L_\varphi^2}=\frac{2n\pi}{L_T^2}}
\end{eqnarray}
The validity of this relation is the same as for eqs.~(\ref{WL},\ref{WL2})~: the system
should be such that it is meaningful to average uniformly the nonlocal
conductivity $\sigma(r,r')$ to get the local conductivity
$\sigma=\int\frac{\D{r}\D{r'}}{\mathrm{Vol}}\sigma(r,r')$. A similar discussion
has been proposed to relate WL and conductivity fluctuations (see appendix E
of Ref.~\cite{TexMon05b}).

Our starting point (\ref{formula51}) is a formulation in the Fourier space,
what implicitly assumes translation invariance. Whereas this assumption seems
reasonable for a large regular network such as the square network studied in
this article, its validity is not clear for networks of arbitrary topology,
what would need further developments.

We have computed the AA correction in a large square network and shown that
the result interpolates between the 1d, eq.~(\ref{see1dlimit}), and a 2d
result, eq.~(\ref{see2dlimit}). Interestingly, the 2d limit in a network
involves a 1d constant~$\lambda^\mathrm{network}_\sigma\simeq4-\frac32F$,
what is confirmed by experiments, as discussed in section~\ref{sec:experiment}.

The interest of the network compared to the plane is to control the constant
$C_\mathrm{ee}$ of eq.~(\ref{see2dlimit})~: for a plane, a cutoff must be
introduced in eq.~(\ref{sigmaee}) at short time $t\sim\tau_e$ and the constant
$C_\mathrm{ee}$ is replaced by a number that depends on the precise cutoff
procedure. Experimentally, it would be interesting to observe the crossover
from (\ref{see1dlimit}) to (\ref{see2dlimit}) by varying $L_T/a$. This was not
possible in experiments of Mallet {\it et al}~\cite{MalSamBau07} described in
section~\ref{sec:experiment} because measurements are complicated by the fact
that electron-phonon interaction also brings a temperature-dependent
contribution, $\Delta\sigma_\mathrm{e-ph}$, at high temperature (above few
Kelvins). The conductivity is given by
$
\sigma=
\sigma_0+\Delta\sigma_\mathrm{WL}+\Delta\sigma_\mathrm{ee}+\Delta\sigma_\mathrm{e-ph}
$.
The WL can be suppressed by a magnetic field however the electron-phonon
contribution is difficult to separate from $\Delta\sigma_\mathrm{ee}$.
Therefore the network should be patterned in a way such that the crossover
1d-2d remains below $T\sim1\:$K where $\Delta\sigma_\mathrm{e-ph}$ is
negligible. As an example we consider the silver networks studied in
Ref.~\cite{SchMalMaiTexMonSamBau07} for which $L_T=0.27\times\:T^{-1/2}$
($L_T$ in $\mu$m and $T$ in K).
In order to see clearly the 1d and the 2d regimes it would be convenient to study
two networks with different lattice spacings. If temperature is
constrained by $10\:$mK$<T<1\:$K, for $a=0.5\:\mu$m we have
$0.5\lesssim{L_T/a}\lesssim5$, which probes the 2d regime over one decade. A
second lattice with $a\sim5\:\mu$m would allow to probe the 1d regime since in
this case~$0.05\lesssim{L_T/a}\lesssim0.5$.

\section*{Acknowledgements}

We have benefitted from stimulating discussions with Christopher B\"auerle,
H\'el\`ene Bouchiat, Meydi Ferrier, Fran\c{c}ois Mallet, Laurent Saminadayar
and F\'elicien Schopfer.

\vfill


\begin{thebibliography}{10}
\bibitem{footnote1}
  It is worth mentioning that a similar effect exists at higher temperature,
  in the ballistic regime ($\tau_\mathrm{ee}\ll\tau_e$)~;
  Ref.~\cite{ZalNarAle01} provides a nice review on this point and describes
  the crossover between the two regimes.
\bibitem{footnote2}  
  Pascaud \& Montambaux have rather considered thermodynamic properties.  The
  nonlocal effects in networks have been further investigated in
  Ref.~\cite{TexMon04}.
\bibitem{footnote3}
  The formula (5.1) of Ref.~\cite{AltAro85} has the wrong sign.
\bibitem{footnote4}
  This interaction assumes that the screening length is smaller than the
  transverse size of the wire.
\end{thebibliography}

\end{document}